\documentclass[a4paper,11pt,reqno]{article}

\usepackage{a4wide}
\usepackage{amsmath,amsfonts,amssymb}
\usepackage[english]{babel}
\usepackage{soul}
\usepackage{nicefrac}
\usepackage[mathscr]{euscript}
\usepackage{setspace}
\usepackage{datetime}
\usepackage[sort&compress,merge,numbers]{natbib}
\usepackage{comment}

\usepackage{mathrsfs}
\usepackage[T1]{fontenc}
\usepackage{mathpazo}
\usepackage{epsfig}

\usepackage[breaklinks=true]{hyperref}

\numberwithin{equation}{section}

\hypersetup{
			colorlinks=true,
			urlcolor=blue,
			citecolor=magenta,
			linkcolor=blue,
     }

\let\oldsqrt\sqrt
\def\sqrt{\mathpalette\DHLhksqrt}

\def\crbig{\\\noalign{\vspace {3mm}}}

\def\DHLhksqrt#1#2{%
\setbox0=\hbox{$#1\oldsqrt{#2\,}$}\dimen0=\ht0
\advance\dimen0-0.2\ht0
\setbox2=\hbox{\vrule height\ht0 depth -\dimen0}%
{\box0\lower0.4pt\box2}}

\def\beq{\begin{equation}}
\def\eeq{\end{equation}}

\def\ov{\overline}

\author{
  \begin{minipage}{.97\linewidth}
    \vspace{1cm}
       \begin{center}
      \begin{small}
        \textbf{Ignatios Antoniadis},$^{1,2}$
      \textbf{Jean-Pierre Derendinger},$^2$\\
     \textbf{P. Marios Petropoulos}$^{3,1}$ and 
      \textbf{Konstantinos Siampos}$^2$
              \end{small}
    \end{center}
    \vspace{0.5cm}
    \hspace{2.4cm}\begin{minipage}{.7\linewidth}
\begin{center}     {\it \begin{footnotesize}
\hbox{\kern-3.cm\vbox{\vskip0cm
 \begin{itemize}
               \item[$^1$] Laboratoire de Physique Th\'eorique et Hautes Energies,\\ 
        Sorbonne Universit\'es, CNRS UMR 7589,\\ 
        UPMC Paris 6, \\
        4 place Jussieu, 75005 Paris, France
\vskip0.29cm
      \end{itemize}}
\kern-3cm\vbox{
\begin{itemize}
 \item[$^2$] Albert Einstein Center for Fundamental Physics,\\
Institute for Theoretical Physics,\\ 
Bern University,\\
Sidlerstrasse 5, 3012 Bern, Switzerland
      \end{itemize}\vskip0.05cm
}}
     \end{footnotesize}}
\end{center}
    \end{minipage}
    \vspace{0.5cm}\begin{minipage}{.7\linewidth}
\begin{center}     
{\it \begin{footnotesize}
\hbox{\kern4.cm\vbox{\vskip0cm
 \begin{itemize}
               \item[$^3$] Centre de Physique Th\'eorique,\\ 
        Ecole Polytechnique, CNRS UMR 7644,\\ 
        Universit\'e Paris-Saclay,\\
        91128 Palaiseau Cedex, France
\vskip0.29cm
      \end{itemize}}
}
     \end{footnotesize}}
\end{center}
     \end{minipage}
  \end{minipage}
}
\date{}

\title{\vspace{3.5cm}
 \boldmath \begin{Large}
    \textbf{Heisenberg symmetry and hypermultiplet manifolds}
  \end{Large} \unboldmath
}

\begin{document}

\begin{titlepage}
\maketitle
\thispagestyle{empty}

 \vspace{-14.5cm}
  \begin{flushright}
  CPHT-040.1015
    \end{flushright}
 \vspace{12cm}

\begin{center}
\textsc{Abstract}\\  
\vspace{1cm}	
\begin{minipage}{1.0\linewidth}

We study the emergence of Heisenberg (Bianchi II) algebra in hyper-K\"ahler and quaternionic  spaces.
This is motivated by the r\^ole these spaces with this symmetry play in $\mathcal{N}=2$ hypermultiplet scalar manifolds. 
We show how to construct related pairs of hyper-K\"ahler and quaternionic  spaces under general symmetry assumptions, the 
former being a zooming-in limit of the latter at vanishing scalar curvature.  We further apply this method for the 
two hyper-K\"ahler spaces with Heisenberg algebra, which is reduced to $U(1)\times U(1)$ at the quaternionic level. 
We also show that no quaternionic spaces exist with a strict Heisenberg symmetry -- as opposed to $\text{Heisenberg} 
\ltimes U(1)$. We finally discuss the realization of the latter by gauging appropriate $Sp(2,4)$ generators in 
$\mathcal{N}=2$ conformal supergravity.

\end{minipage}
\end{center}


\end{titlepage}

\onehalfspace

\vspace{-1cm}
\begingroup
\hypersetup{linkcolor=black}
\boldmath
\tableofcontents
\unboldmath
\endgroup
\noindent\rule{\textwidth}{0.6pt}

\section*{Introduction}
\addcontentsline{toc}{section}{Introduction}

In string theory, the Heisenberg algebra appears within the universal hypermultiplet of type IIA compactification
\cite{Fer}. The dilaton is contained in the scalar manifold, which is a four-dimensional quaternionic space 
\cite{BW}.\footnote{In the literature, manifolds with holonomy contained in $Sp(2)\times Sp(2n)$ and non-zero 
Ricci curvature are actually called quaternion-K\"ahler. In the four-dimensional case of interest here, they are
Weyl-self-dual Einstein spaces.}
At tree level,  the latter is $\widetilde{\mathbb{C}P}_2$ equipped with the K\"ahler (non-compact) Fubini--Study metric
 of $SU(1,2) /U(2)$. Perturbative corrections break the large isometry group of this space to the Heisenberg group, 
generated by three shifts (NSNS axion and RR scalar) \cite{oneloop}.  More precisely, it was observed that the residual 
symmetry is rather $\text{Heisenberg} \ltimes U(1)$, and that this symmetry uniquely determines the quaternionic space. 
Non-perturbative corrections further break the Heisenberg symmetry down to $U(1)\times U(1)$ \cite{Alexandrov:2008nk}
(or generically to a discrete subgroup of the Heisenberg group -- see for example \cite{Bao:2009fg}). The corresponding scalar manifold is thus a quaternionic space with two commuting Killing vectors. Metrics on
these manifolds have been characterized by Calderbank and Pedersen \cite{Pedersen}.

In the above framework, supersymmetry is locally realized and the scalar curvature of the quaternionic space 
is directly proportional to the gravitational constant $k^2 =8\pi  M_{\text{Planck}}^{-2}$ \cite{BW}.
For hypermultiplets of global ${\cal N}=2$, 
the relevant sigma-model target spaces are hyper-K\"ahler \cite{AlvarezGaume:1981hm}. These K\"ahler
spaces are Ricci-flat and, in the four-dimensional case, Riemann self-dual \emph{i.e.}~they are gravitational instantons.
There exists then plausibly a low-energy decoupling limit of gravity $M_{\text{Planck}}\to\infty$, which
deforms the quaternionic geometry into a hyper-K\"ahler limit. Since any hyper-K\"ahler
manifold can be coupled to supergravity in a quaternionic manifold, this limiting process must smoothly interpolate between both
geometries, and its description requires care. 
It implies to simultaneously ``zooming-in'' with appropriate $k$ factors in order to recover 
non-trivial hyper-K\"ahler geometries~\cite{AADT}. This procedure 
has been demonstrated for specific cases, involving the quaternionic quotient method \cite{Galicki:1985qv,Galicki:1987jz}.

Although, as pointed out previously, the Heisenberg algebra is uniquely realized at the quaternionic level as 
$\text{Heisenberg} \ltimes U(1)$, two distinct hyper-K\"ahler spaces exist with Bianchi II symmetry, realized either as $\text{Heisenberg} \ltimes U(1)$ (biaxial), or as strict $\text{Heisenberg}$ (trixial) \cite{BEPS}. The former corresponds indeed to the infinite-$M_{\text{Planck}}$ limit of the quaternionic space sharing its isometry and describing the string perturbative corrections to the hypermultiplet manifold, whereas nothing is known about the latter. 

The purpose of the present note is to elaborate on the hyper-K\"ahler space with strict  Heisenberg isometry. 
This raises a number of interesting questions, some of which stand beyond Heisenberg symmetry: 
\begin{itemize}
\item 
Under which conditions a hyper-K\"ahler space of a prescribed isometry can give rise to a quaternionic ascendent with the same or less symmetry?
\item Conversely, what is the general limiting procedure for reaching smoothly a non-trivial hyper-K\"ahler space from a given quaternionic one? 
\end{itemize}
Examples are known, where  a quaternionic space is constructed starting from a four-dimen\-sio\-nal hyper-K\"ahler geometry via an 
eight-dimensional hyper-K\"ahler cone \cite{deWit:1999fp,Anguelova:2004sj}. 
An interesting relationship can be further settled amongst quaternionic spaces with an isometry and  hyper-K\"ahler spaces with a 
rotational symmetry, equipped with a hyper-holomorphic connection (\emph{i.e.}
whose curvature is $(1,1)$ with respect to all complex structures in the hyper-K\"ahler family). This was developed 
in \cite{Haydys,Hitchin2012}  from a mathematical point of view, and in \cite{Alexandrov:2011ac} in a more physical framework.
We will here provide an alternative and systematic algebraic procedure for a direct uplift, together with a scaling descent method. 
These general tools make it possible for investigating the more specific case of Heisenberg symmetry, reaching the following conclusions:
\begin{itemize}
\item There are no quaternionic spaces with triaxial Heisenberg isometry.
\item The hyper-K\"ahler space with strict Heisenberg symmetry admits a quaternionic ascendent with $U(1)\times U(1)$ isometry.
\item The hyper-K\"ahler space with  $\text{Heisenberg} \ltimes U(1)$ symmetry admits yet another quaternionic ascendent with $U(1)\times U(1)$ isometry, besides the known one with $\text{Heisenberg} \ltimes U(1)$ symmetry of which it is the low-energy limit. 
\item The known quaternionic space with $\text{Heisenberg} \ltimes U(1)$ symmetry is the extended symmetry point of an $Sp(2,4)$ gauging producing a family of Calderbank--Pedersen spaces.
\end{itemize}
In contrast with the biaxial case which captures well-identified perturbative string contributions \cite{oneloop}, 
neither for the triaxial-Heisenberg hyper-K\"ahler, nor for the two quaternionic ascendents with  $U(1)\times U(1)$ symmetry is such an interpretation available, though, and this issue is left for future work.

In the following we will first summarize the results regarding the realization of Heisenberg symmetry in 
hyper-K\"ahler spaces, Sec. \ref{hyper}. We will then move to quaternionic spaces in  Sec. \ref{quater}, and discuss the realization of Heisenberg symmetry, in particular the obstruction to a strict Heisenberg isometry group. The general procedure for taking the gravity-decoupling limit will also be presented, along with the systematic method for building up quaternionic ascendents, based on the existence of a Boyer--Finley 
 field representation in quaternionic and hyper-K\"ahler spaces. As mentioned earlier, there are alternative methods for scanning these spaces, based on gaugings, which will be exposed in Sec. \ref{gaug}. Two appendices complete the technical details, in particular regarding the gauging procedure.  

\section{Hyper-K\"ahler spaces with Heisenberg symmetry}\label{hyper}

\subsection{Translational vs rotational isometries} 

A four-dimensional hyper-K\"ahler space is Ricci-flat with (anti-)self-dual Riemann tensor:
\begin{equation}
\label{self.duality}
R_{\kappa\lambda\mu\nu}=\pm\frac12\,\varepsilon_{\kappa\lambda}{}^{\rho\sigma}\,R_{\rho\sigma\mu\nu}\quad
\text{with}\quad\varepsilon_{\kappa\lambda\mu\nu}=\sqrt{\det g}\, \epsilon_{\kappa\lambda\mu\nu}\,,\quad \epsilon_{0123}=1\,,
\end{equation}
($+$ corresponds to self-duality). In the presence of an isometry generated by a Killing vector $\xi=\xi^\mu\partial_\mu$, using the Bianchi identity for the Riemann tensor, it is known that
\begin{equation}
 \nabla_\mu\nabla_\lambda\xi_\kappa=R_{\kappa\lambda\mu\nu}\xi^\nu\,,
\end{equation}
and consequently we can prove that 
\eqref{self.duality} is equivalent to
\begin{equation}
\nabla_\mu \left(\nabla_\lambda \xi_\kappa\mp\frac{1}{2}\varepsilon_{\lambda\kappa}{}^{\sigma\rho}\,\nabla_\sigma \xi_\rho\right)=0\,.
\end{equation} 
If
\begin{equation}
\nabla_\lambda\xi_\kappa=\pm\frac12\,\varepsilon_{\lambda\kappa}{}^{\sigma\rho}\,\nabla_\sigma\xi_\rho\,,
\end{equation}
the Killing $\xi$ is a translational vector; it is  otherwise rotational. 

Using the Killing vector at hand, we can adapt a coordinate $\tau$ to it,  $\xi=\partial_\tau$,
and write the metric as a fiber
along this isometry:
\begin{equation}
\label{ds}
\text{d}s^2=\frac{1}{V}\left(\text{d}\tau+\omega_i\text{d}x^i\right)^2+V\text{d}\ell^2
\end{equation}
with
\begin{equation}
\label{dl}
\text{d}\ell^2=\gamma_{ij}\,\text{d}x^i\text{d}x^j\,,\qquad i=1,2,3\,,
\end{equation}
where we note the gauge invariance $\delta\tau = f(\vec x)$, $\delta\vec \omega = -\vec\nabla f(\vec x)$.

When $\partial_\tau$ is a translational Killing vector, one is allowed to use the Gibbons--Hawking frame \cite{Gibbons:1979zt},
\begin{equation}
\text{d}V=\pm\star_\gamma\text{d}\omega\,,\qquad\gamma_{ij}=\delta_{ij}\,,
\end{equation}
or in everyday's language
\begin{equation}
\vec\nabla V = \pm \vec\nabla\wedge\vec \omega,
\end{equation}
whose compatibility yields the condition
\begin{equation}
\label{Lap1}
\vec\nabla^2 \, V=0. 
\end{equation}

When $\partial_\tau$ is rotational, we can rewrite the metric in the Boyer--Finley frame \cite{Boyer},
where expression (\ref{ds}) now holds with
\begin{equation}
\label{Toda.frame}
\text{d}\ell^2=\text{d}Z^2+\text{e}^{\Psi} \left(\text{d}X^2+\text{d}Y^2\right),
\end{equation}
and $V, \omega$ are now given by
\begin{equation}
\label{Vom}
V=\frac12\,\partial_Z\Psi\,,\qquad \omega_X=\frac12\,\partial_Y\Psi\,,\qquad
\omega_Y=-\frac12\,\partial_X\Psi\,.
\end{equation}
The third component $w_Z$ vanishes by a gauge choice of the coordinate $\tau$.
Now the (anti-)self-duality condition requires vanishing of the Laplacian over $\text{d}\ell^2$ in \eqref{Toda.frame}:
\begin{equation}
\label{eqpsi}
\Delta\,\Psi=0\Longleftrightarrow\left(\partial^2_X+\partial^2_Y\right)\Psi+\partial^2_Z\,\text{e}^\Psi=0\,.
\end{equation}
This equation is known as the continual Toda equation, found in the context of continuum Lie algebras \cite{Saveliev}.
Notice that V satisfies
\begin{equation}
\left(\partial_X^2 + \partial_Y^2\right)V + \partial_Z^2\left(V \text{e}^\Psi \right)=0
\end{equation}
instead of condition (\ref{Lap1}).

Finally we note that the translational case has an alternative formulation. Start with a function $\Psi$ solution of the (flat-space) 
Laplace equation $\vec\nabla^2\Psi=\left(\partial_X^2+\partial_Y^2+\partial_Z^2\right)\Psi=0$, and write
\eqref{Vom},
\begin{equation}
V = \frac12\partial_Z\Psi\, , \qquad \omega_X=\frac12\partial_Y\Psi\, , \qquad \omega_Y= -\frac12\partial_X\Psi\, ,
\nonumber
\end{equation}
in the gauge $\omega_Z=0$. Therefore the function $\Psi$ generates the metric without appearing explicitly in it.

\subsection{The two geometries with Heisenberg symmetry}\label{heishyp}

In a systematic investigation of four-dimensional gravitational instantons with Bianchi isometry group, the general solution for Bianchi-II type (Heisenberg symmetry) was found in Ref.~\cite{BEPS}. The corresponding Riemann self-dual metrics read:
\begin{equation}
\label{hyp1}
\text{d}s^2 = \frac{1}{t} (\sigma^1)^2 
+ t \left[(\sigma^2)^2+\text{e}^{2\varepsilon t} \left(\text{d}t^2 + (\sigma^3)^2 \right) \right]\,,
\end{equation}
with $\varepsilon\geqslant0$ a continuous parameter. When non-zero, this parameter can be reabsorbed in a coordinate redefinition, hence this family contains only two members. Here
\begin{equation}
\label{hyp2}
\text{d}\sigma^1 = \sigma^2\wedge\sigma^3, \qquad \text{d}\sigma^2=\text{d}\sigma^3=0
\end{equation}
are the Maurer--Cartan left-invariant forms of Bianchi II. These are here realized as
\begin{equation}
\label{hyp3}
\sigma^1 = \text{d}z + x\text{d}y\,, \qquad
\sigma^2 = \text{d}x\,, \qquad
\sigma^3 = \text{d}y\,.
\end{equation}
They are invariant under the Killing fields 
\begin{equation}
\mathscr{X}=\partial_x-y \partial_z\,, \qquad
\mathscr{Y}=\partial_y\,,\qquad
\mathscr{Z}=\partial_z\,, 
\end{equation}
which obey the Heisenberg algebra:
\begin{equation}
\left[\mathscr{X}, \mathscr{Y}\right]=\mathscr{Z},\qquad
\left[\mathscr{Z}, \mathscr{X}\right]=\left[\mathscr{Z}, \mathscr{Y}\right]=0.
\end{equation}
While $\mathscr{X}$ and $\mathscr{Z}$ are always translational, $\mathscr{Y}$ is rotational when $\varepsilon>0$, or translational if $\varepsilon=0$.
Besides the generators of the Heisenberg algebra, the vector field
\begin{equation}
\label{xi}
\mathscr{M}=y \partial_x-x \partial_y+\frac{1}{2}\left(x^2-y^2\right) \partial_z=
y\,\mathscr{X}-x\,\mathscr{Y} +\frac{1}{2}\left(x^2+y^2\right)\mathscr{Z}
\end{equation}
turns out to play a r\^ole. It has the following commutation relations with $\mathscr{X},\mathscr{Y},\mathscr{Z}$:
\begin{equation}
\label{Heis}
\left[\mathscr{M}, \mathscr{X}\right]=\mathscr{Y}\,,
\qquad\left[\mathscr{M}, \mathscr{Y}\right]=-\mathscr{X}\,, \qquad\left[\mathscr{M}, \mathscr{Z}\right]=0\,.
\end{equation}
These define a semi-direct product of a $U(1)$ with the Heisenberg algebra, $\mathscr{Z}$ being the center of the 
resulting four-dimensional algebra.

The hyper-K\"ahler metric \eqref{hyp1} 
\begin{equation}
\label{hyp7}
\text{d}s^2 =\frac{1}{t} (\text{\text{d}}z + x \, \text{\text{d}}y)^2 + t\left[ \text{\text{d}}x^2 
+
 \text{e}^{2\varepsilon t} \left(\text{\text{d}}y^2+ \text{\text{d}}t^2\right)  \right]
 \end{equation}
is invariant under the full Heisenberg algebra. For vanishing $\varepsilon$, it is also invariant under $\mathscr{M}$, which turns out to be rotational. It is common to call \emph{triaxial} the realization of strict Heisenberg isometry as it occurs for $\varepsilon >0$, and \emph{biaxial} the case where it is accompanied with an extra $U(1)$.

\subsection{K\"ahler coordinates}\label{kaehler}

Several K\"ahler coordinate systems are available for the spaces \eqref{hyp7}, providing various realizations  of the 
$\text{Heisenberg} \ltimes U(1)$ algebra. We can use for example a set of K\"ahler coordinates $\Phi$ and $T$ defined as
\begin{equation}
\label{Kceps0}
\Phi={t+iy}\,,\quad T=-tx+iz\,.
\end{equation}
and the K\"ahler potential is given by
\begin{equation}
\label{Kis}
K=\frac{\left(T+\ov T\right)^2}{\Phi+\ov\Phi}
+\frac{1}{2\varepsilon^3}\,\left[\varepsilon\left(\Phi+\ov\Phi\right)-2\right]\text{e}^{\varepsilon\,\left(\Phi+\ov\Phi\right)}\, .
\end{equation}
In these coordinates, the 
Heisenberg algebra is realized as:
\begin{equation}
\label{real1}
\mathscr{X}=-\Phi\,\partial_T-\ov\Phi\,\partial_{\ov T}\,,\quad 
\mathscr{Y}=i\left(\partial_\Phi-\partial_{\ov\Phi}\right)\,,\quad
\mathscr{Z}=i(\partial_T-\partial_{\ov T})\,,
\end{equation}
under which the K\"ahler potential is invariant up to the following K\"ahler transformation:
\begin{equation}
\label{Kt}
\mathscr{X}(K)=-2(T+\ov T)\,,\quad \mathscr{Y}(K)=\mathscr{Z}(K)=0\,.
\end{equation}
The extra $U(1)$ generator \eqref{xi} reads now:
 \begin{equation}
\label{Meps1}
\mathscr{M}=\frac{1}{2i}\left(\Phi-\ov\Phi\right)\mathscr{X}
+\frac{T+\ov T}{\Phi+\ov\Phi}\,\mathscr{Y}+
\frac12\,\left(\left(\frac{T+\ov T}{\Phi+\ov\Phi}\right)^2-\frac14\,\left(\Phi-\ov\Phi\right)^2\right)\,\mathscr{Z}\,.
\end{equation}
It completes the $\text{Heisenberg} \ltimes U(1)$ algebra.

Note that $\mathscr{M}$ is a linear combination of the Heisenberg group generators 
$\mathscr{X}$, $\mathscr{Y}$ and $\mathscr{Z}$ with field-dependent non-holomorphic 
coefficients, and thus does not correspond to a holomorphic transformation in 
these coordinates, in contrast to $\mathscr{X}$, $\mathscr{Y}$, $\mathscr{Z}$. 
Consequently, the variation of the line element cannot be derived from the variation 
of the K\"ahler potential, but instead by direct computation of its Lie derivative:
\beq
\begin{array}{rcl}
\mathcal{L}_{\mathscr M} \, {\rm d}s^2 &=& \displaystyle 2t \left[ 1 - \text{e}^{2\varepsilon t} \right]  {\rm d}x\,{\rm d}y
\crbig 
&=& \displaystyle
\frac{i}{2}\left[ 1 - \text{e}^{\varepsilon(\Phi+\ov\Phi)} \right]  \left( {\rm d}(T+\ov T) - {T+\ov T \over \Phi+\ov\Phi}\,
{\rm d} (\Phi+\ov\Phi) \right) {\rm d}(\Phi-\ov\Phi) \, .
\end{array}
\eeq
This vanishes only for $\varepsilon = 0$, in which case $\mathscr M$ generates  a symmetry.

Thus, for vanishing $\varepsilon$, the vector $\mathscr{M}$ is the generator of an extra isometry, promoting the 
symmetry to $\text{Heisenberg} \ltimes U(1)$
with K\"ahler potential given by the finite part of \eqref{Kis} in the limit $\varepsilon \to 0$ (the divergent terms are harmonic functions and play no r\^ole as they can be reabsorbed by 
K\"ahler transformations):
\begin{equation}
K=\frac{(T+\ov T)^2}{\Phi+\ov\Phi}+\frac{(\Phi+\ov\Phi)^3}{12}\,.
\end{equation}
In the $\varepsilon = 0$ case at hand, the isometry generator $\mathscr{M}$ acts as a rotation in the  $(x,y)$-plane. As already mentioned, 
this action is non-holomorphic on the K\"ahler coordinates $(T,\Phi)$, 
but  alternative sets of complex fields exist, in which all isometries 
are holomorphically implemented. We may choose for example:
\begin{equation}
\label{hk8}
\Psi = x+iy\, , \quad U=\frac{1}{2}\left(t^2-x^2\right)+iz \,,
\end{equation}
in which case the $\varepsilon=0$ K\"ahler potential is given by
\begin{equation}
\label{HK10}
K = \frac{4}{3}\,  Q^{\nicefrac{3}{2}}\, , 
\end{equation}
where 
\begin{equation}
Q = U+\ov U + \frac{1}{4}(\Psi+\ov\Psi)^2\, .
\end{equation}
The generators of the four isometries become
\begin{equation}
\begin{split}
&\mathscr{X}=-\Psi\partial_U-\ov\Psi\partial_{\ov U}+\partial_\Psi+\partial_{\ov\Psi}\,,
\quad \mathscr{Y}=i\left(\partial_\Psi-\partial_{\ov\Psi}\right),\quad \mathscr{Z}=i\left(\partial_U-\partial_{\ov U}\right),\\
&\mathscr{M}=-i\left(\Psi\partial_\Psi-\ov\Psi\partial_{\ov\Psi}\right)+\frac{i}{2}\left(\Psi^2\partial_U-\ov\Psi^2\partial_{\ov U}\right)\,,
\end{split}
\end{equation}
and $Q$ is now invariant under all transformations. These are the coordinates used in Ref. \cite{AADT} and the relations between the three coordinate sets introduced here are as follows:
\begin{equation}
t=\text{Re}\Phi=\sqrt{Q}\,,\quad x=-\frac{\text{Re}\,T}{\text{Re}\,\Phi}=\text{Re}\,\Psi\,,\quad
y=\text{Im}\,\Phi=\text{Im}\Psi\,,\quad z=\text{Im}\,T=\text{Im}\,U\,.
\end{equation}

\section{Quaternionic uplifts}\label{quater}

\subsection{Przanowski--Tod and Calderbank--Pedersen spaces}

In the framework of ${\cal N}=2$ supergravity, we are here interested in four-dimensional quaternionic  spaces with at least one shift symmetry, and these are part  of a wide web of geometries with remarkable properties.
A four-dimensional quaternionic space is an Einstein space with $R=-12k^2$ ($k$ defining an overall scale in Planck units) and  self-dual  Weyl tensor. This space is always determined by a solution of the continual Toda equation \cite{prz0, prz1, Tod, Tod:2006wj}. If $\partial_\tau$ is the Killing field generating the shift symmetry, the metric on this space reads:
\begin{equation}
\label{prztod}
\text{d}s^2=\frac{1}{Z^2}\left(\frac{1}{U}(\text{d}\tau+\text{A})^2+U\text{d}\ell^2\right),
\end{equation}
with  $\text{d}\ell^2$ of the form \eqref{Toda.frame}, and $\Psi$ satisfying the continual Toda equation \eqref{eqpsi},
\begin{equation*}
\left(\partial^2_X+\partial^2_Y\right)\Psi+\partial^2_Z\text{e}^ \Psi=0\,.
\end{equation*}
The form $\text{A}$ obeys
\begin{equation}
\label{Aform}
\text{dA}=\partial_X U \, \text{d}Y\wedge \text{d}Z+\partial_Y U \, \text{d}Z\wedge \text{d}X+\partial_Z\left(U\, \text{e}^\Psi\right)\text{d}X\wedge \text{d}Y
\end{equation}
with integrability condition:
\begin{equation}
\label{eqU}
\left(\partial_X^2+\partial_Y^2\right)U+\partial_Z^2\left(U\text{e}^\Psi\right)=0,
\end{equation}
whereas $U$ and $\Psi$ are constrained by
\begin{equation}
\label{quatreq}
2k^2 U=2-Z\partial_Z \Psi.
\end{equation}
Note that Eq. \eqref{eqU}, known as linearized Toda equation, is compatible with \eqref{eqpsi} and
\eqref{quatreq}.
The corresponding quaternionic space is commonly known as 
Przanowski--Tod.

The Calderbank--Pedersen metrics studied in Ref. \cite{Pedersen} are the most general quaternionic spaces with two commuting isometries (as they appear e.g. in the Heisenberg algebra \eqref{Heis}). Hence, they belong to the Przanowski--Tod family. The generic  Calderbank--Pedersen metric is expressed in terms of a function $F(\rho, \eta)$, where $\rho$ and $\eta$ are two coordinates and
the other two, $\tau$ and $\psi$, support the two  Killing vectors $\partial_\tau,\partial_\psi$. The function $F$ is an eigenfunction of the Laplacian operator on the hyperbolic two-plane with metric $\frac{\text{d}\rho^2+\text{d}\eta^2}{\rho^2}$, for eigenvalue $\nicefrac{3}{4}$. Trading $F$ for $G=\sqrt{\rho} F$ in the original  Calderbank--Pedersen expression, one obtains: 
\begin{equation}
\label{CPprztod}
\text{d}s^2=\frac{1}{G^2}\left(\frac{1}{U}(\text{d}\tau+\text{A})^2+U\text{d}\ell^2\right),
\end{equation}
with\footnote{Indices indicate derivatives with respect to $\rho$, or $\eta$.}
\begin{equation}
\label{CPprztod3d}
 \text{d}\ell^2=\left(G_\rho^2+G_\eta^2\right)\left(\text{d}\rho^2+\text{d}\eta^2\right)+\rho^2\text{d}\psi^2 ,\end{equation}
and
\begin{equation}
\label{CPgeneral1}
A=\frac{1}{k^2}\left(\eta-\frac{GG_\eta}{G_\rho^2+G_\eta^2}\right)\text{d}\psi,\quad
U=\frac{1}{k^2}\left(1-\frac{1}{\rho} \frac{GG_\rho}{G_\rho^2+G_\eta^2} \right),
\end{equation}
where
\begin{equation}
\rho\left(G_{\rho\rho}+G_{\eta\eta}\right)=G_\rho\,.
\end{equation}
Some efforts are needed to further turn the Calderbank--Pedersen metric \eqref{CPprztod} into the Prza\-now\-ski--Tod form \eqref{prztod}, by setting $Z=G(\rho,\eta)$ and expressing $X(\rho,\eta), Y(\rho,\eta)$ and $\Psi(\rho,\eta)$ in terms of $G$. The interested reader can find details in Refs. \cite{Pedersen,Casteill:2001zk}.

\boldmath
\subsection{From Boyer--Finley to Przanowski--Tod and back}\label{back}
\unboldmath

The key observation regarding hyper-K\"ahler spaces with a rotational Killing vector, on the one hand, and quaternionic spaces with a symmetry, on the other, is that they share the Boyer--Finley frame \eqref{Toda.frame} and Toda equation \eqref{eqpsi}. In other words, a solution $\Psi(X,Y,Z)$ of Toda equation \eqref{eqpsi} can either produce a hyper-K\"ahler space in  Boyer--Finley form \eqref{ds}, when combined with 
\eqref{Vom}, or a quaternionic space in Przanowski--Tod form \eqref{prztod}, when combined with 
\eqref{quatreq}. Both have at least one isometry generated by $\partial_\tau$. As we will see, extra isometries, if present in one,  may or may not be realized in the other. 

The relationship between the pair of spaces built around one solution of Toda equation is even more intimate. Indeed, the hyper-K\"ahler member turns out to be the $k \to 0$ limit of the quaternionic one, the limit being taken in an appropriate zoom-in manner for avoiding the trivialization of the geometry into flat space. For that, consider the following transformation:
\begin{equation}
{Z\mapsto Z-\delta}\,,\quad U= \delta^2V\,,\quad \tau\mapsto \delta^2\tau\,,\quad \text{A}= \delta^2\omega\,.
\end{equation}
Performed on the Przanowski--Tod metric \eqref{prztod}, on the form \eqref{Aform} and on the constraint equation \eqref{quatreq}, these read: 
\begin{eqnarray}
\text{d}s^2&=&{\frac{\delta^2}{(Z-\delta)^2}}\left(\frac{1}{V}\left(\text{d}\tau+\omega\right)^2+V\left[\text{d}Z^2+\text{e}^\Psi\left(\text{d}X^2+\text{d}Y^2\right)\right]\right)\,,\\
\text{d}\omega&=&\partial_X V \, \text{d}Y\wedge \text{d}Z+\partial_Y V \, \text{d}Z\wedge \text{d}X+\partial_Z\left(V\, \text{e}^\Psi\right)\text{d}X\wedge \text{d}Y\, , \\
V&=&{\frac{1}{2\delta\,k^2}\,\partial_Z\Psi+\frac{1}{2\delta^2\,k^2}\left(2-Z\partial_Z\Psi\right)}\,.
\end{eqnarray}
In the double-scaling limit
\begin{equation}
k \to0\,,\quad \delta\to\infty\,,\quad k^2\,\delta=1\,,
\end{equation}
we recover a Ricci-flat (anti-)self-dual instanton \eqref{ds} in Boyer--Finley frame \eqref{Toda.frame} with $V, \omega$ satisfying \eqref{Vom}. Whenever the quaternionic space is the manifold of a hypermultiplet coupled to ${\cal N}=2$ supergravity, the gravitational constant is $k^2=8\pi M_{\text{Planck}}^{-2}$ \cite{BW}, and the limit $k\to0$ corresponds to a hypermultiplet of global ${\cal N}=2$.

\boldmath
\subsection{The Heisenberg algebra in quaternionic spaces} \label{genquater}
\unboldmath

\boldmath
\subsubsection*{An obstruction for quaternionic spaces with triaxial Heisenberg symmetry}
\unboldmath

The two hyper-K\"ahler spaces with Heisenberg algebra discussed earlier in Sec.~\ref{heishyp}, may be uplifted to quaternionic. Indeed, in both cases, a rotational  Killing vector exists, and a Boyer--Finley frame can be exhibited, with a solution to Toda equation.  Before analyzing these two specific spaces, we would like to demonstrate a general property, which will be illustrated afterwards: \emph{no four-dimensional quaternionic space exists with strict Heisenberg isometry}. 

There are at least two ways to prove this statement. 
Firstly, using isomonodromic deformations, a method developed by 
Hitchin \cite{Hitchin} and Tod \cite{Tod.Einstein}; secondly using foliations with Heisenberg isometry.
Here we choose the most economical one with the tools at hand, which is the second. 

A general four-dimensional geometry with Heisenberg symmetry can be realized as a Bianchi-II foliation. 
We consider foliations of the type:  
 \begin{equation}\label{metans}
\mathrm{d}s^2 
= a^2b^2c^2\, 
\mathrm{d}t^2
    +
   a^2\, \big(\sigma^1\big)^2+
    b^2\left(\sigma^2\right)^2+
    c^2\left(\sigma^3\right)^2
    \end{equation}
with $\sigma^i$ the Bianchi II Maurer--Cartan forms, given in \eqref{hyp3} and obeying \eqref{hyp2}. Here $a,b,c$ are functions of $t$ and fully characterize the geometry, which is by construction invariant under the Heisenberg algebra. 

Two remarks are in order. Firstly, we might have chosen $g_{ij}(t)$ instead of  $\text{diag} (a,b,c)$. For unimodular Bianchi groups, however, such a non-diagonal form can always be brought into a diagonal one, and our choice is not restrictive (for a systematic analysis of this issue, see \cite{BEPS}). Secondly, the metric under consideration has strict Heisenberg symmetry, as long as $b$ and $c$ are not proportional to each other. When $b\propto c$, an extra $U(1)$ appears, generated by $\mathscr{M}$ given in \eqref{xi} (up to appropriate rescaling of the coordinates in order to reabsorb the constant $\nicefrac{b}{c}$).

We now impose that \eqref{metans} satisfies Einstein's equations:
\begin{equation}
R_{\alpha\beta}=\frac{R}{4}\,g_{\alpha\beta}\,,
\end{equation}
where the scalar curvature is constant: $R=-12k^2 $. We find
\begin{equation}
\label{jhdfsjlklkjsd}
\dot a=-\frac12a^3+abc\lambda\,,\quad \dot b=\frac12\,a^2b+abc\mu\,,\quad  \dot c=\frac12\,a^2c+abc\nu\,,
\end{equation}
(the dot stands for the derivative with respect to $t$) where $\lambda(t), \mu(t), \nu(t)$
are first integrals obeying
\begin{equation}
\label{jhdfsjlklkjsd2}
\dot\lambda=a^2(\lambda+\mu\nu)\,,\quad\dot\mu=b^2\lambda\nu\,,\quad \dot\nu=c^2\lambda\mu\,.
\end{equation}
The constant scalar curvature further imposes 
\begin{equation}
\label{Rk}
a(\lambda+\mu\nu)+b\lambda\nu+c\lambda\mu=3k^2abc\,.
\end{equation}

The requirement of Weyl self-duality is the next step:
\begin{equation}
W_{\kappa\lambda\mu\nu}=\frac12\,\varepsilon_{\kappa\lambda}{}^{\rho\sigma}\,W_{\rho\sigma\mu\nu}\,.
\end{equation}
Using \eqref{jhdfsjlklkjsd} and \eqref{jhdfsjlklkjsd2}, three distinct cases emerge:
\begin{equation}
\begin{cases}
\label{jdjdjfdjfj}
\lambda=\nu=0\,, \\
\lambda=\mu=0\,,\\
b=a\frac{\lambda+\mu\nu}{\lambda\nu}\ \text{and}\ 
c=a\frac{\lambda+\mu\nu}{\lambda\mu}\,.
  \end{cases}
\end{equation}
Owing to Eq. \eqref{Rk}, the first two cases have vanishing scalar curvature, and correspond therefore to the Ricci-flat self-dual instantons displayed in Eq. \eqref{hyp7}.\footnote{If $\lambda=\mu=0$ then Eqs. \eqref{jhdfsjlklkjsd}, \eqref{jhdfsjlklkjsd2} can be easily integrated with solution
\begin{equation*}
a=t^{\nicefrac{-1}{2}}\,,\quad b=t^{\nicefrac{1}{2}}\,,\quad c=t^{\nicefrac{1}{2}}\,\text{e}^{\varepsilon t}\,,
\end{equation*}
where $\nu$ is a constant parameterized by $\varepsilon$. A similar analysis can be repeated for the other case, $\lambda=\nu=0$.} For the third one in \eqref{jdjdjfdjfj}, clearly
\begin{equation}
\frac{b}{c}=\frac{\mu}{\nu}\, .
\end{equation}
Differentiating the latter with respect to $t$ and using Eqs. \eqref{jhdfsjlklkjsd2}, we obtain
\begin{equation}
\frac{\text{d}}{\text{d}t}\left(\frac{b}{c}\right)=
\lambda\frac{b^2\nu^2-c^2\mu^2}{\nu^2}=0\,.
\end{equation}
The ratio $\nicefrac{b}{c}$ remains thus constant, and the Heisenberg algebra has a biaxial realization in the quaternionic space \eqref{metans} at hand: an extra Killing vector field emerges. 
Picking up for convenience $b=c$ and so $\mu=\nu$, we find the general solution of \eqref{jhdfsjlklkjsd}, \eqref{jhdfsjlklkjsd2}:  
\begin{equation}
\label{abc}
a^2=\frac{8\rho^2}{k^2V_1V_2^2}\,,\quad b^2=\frac{2V_1}{k^2V^2_2}\,,\quad \lambda=-\frac{2}{V_2}\,,\quad \mu=-\frac{2\rho}{V_2}
\end{equation}
with
\begin{equation}
\label{V1V2}
V_1=\rho^2+2\sigma\,,\quad V_2=\rho^2-2\sigma\,,\quad \sigma=\text{constant},
\end{equation}
where $t$ has been traded for a new coordinate $\rho$:
\begin{equation}
\label{trho}
\mathrm{d}t =  \frac{k^2V_2^2}{4\rho}\mathrm{d}\rho\Longrightarrow t=\frac{k^2}{16}\left(\rho^4-8\sigma\rho^2+16\sigma^2\ln\rho\right)
\end{equation}
(up to an irrelevant additive constant reabsorbed in a redefinition of $t$). Trading $(x,y,z)$ 
for $(\eta,\psi,\tau)$, the quaternionic space \eqref{metans} reached with \eqref{abc}, \eqref{V1V2} and \eqref{trho} is 
\begin{equation}
\label{CPM}
\boxed{
\mathrm{d}s^2=\frac{8\rho^2}{k^2V_1V_2^2}\left(\mathrm{d}\tau+\eta\mathrm{d}\psi\right)^2+\frac{2V_1}{k^2V^2_2}\left(\mathrm{d}\rho^2+\mathrm{d}\eta^2+\mathrm{d}\psi^2\right).}
\end{equation}
This is the $\mathcal{N}=2$ hypermultiplet scalar manifold that captures string one-loop perturbative corrections, found in \cite{oneloop}. The metric \eqref{CPM} has well-defined $M_{\text{Planck}}\to\infty$ limit, which coincides with the $\text{Heisenberg}\ltimes U(1)$-symmetric hyper-K\"ahler space \eqref{hyp7} at $\varepsilon = 0$ (see \cite{AADT}). 

In conclusion, the Heisenberg symmetry is always biaxially realized at the quaternionic level, leading to the geometry \eqref{CPM}: no triaxial Heisenberg quaternionic space exists.
We will meet an instance of this obstruction in the following, where the pure Heisenberg symmetry present at the hyper-K\"ahler level is 
actually broken to its $U(1)\times U(1)$ subgroup when moving to the quaternionic. This general conclusion is in agreement with 
the result of Ref.~\cite{Anguelova:2004sj}, where the quaternionic space obtained with Heisenberg 
symmetry has automatically an additional $U(1)$ isometry.

\boldmath
\subsubsection*{The $\varepsilon=0$ uplift}
\unboldmath

The corresponding hyper-K\"ahler metric, in the form  \eqref{hyp7} with $\varepsilon=0$, is Gibbons--Hawking, as the fiber is supported by  $\mathscr{Z}$, which is translational. In this case the realization of the Heisenberg symmetry is biaxial and the rotational Killing vector is the extra generator $\mathscr{M}$. We have to adopt it for the fiber and define $\tau$ such as $\mathscr{M}=\partial_\tau$.
Together with $\tau$, we introduce new coordinates $X,Y,Z$, for which we trade $t,x,y$ and $z$:
\begin{equation}
\label{XYZxyz}
\tau=\arctan \frac{x}{y}\,,\quad
X=z+\frac{xy}{2}\,,\quad Y=\frac{1}{4}\left(x^2+y^2-2t^2
\right), \quad Z=\frac{t}{2}\left(x^2+y^2\right).
\end{equation}
With these new coordinates, the metric \eqref{hyp7} assumes the Boyer--Finley form \eqref{ds}, \eqref{Toda.frame} and \eqref{Vom}, with $\Psi(Y,Z)$ given by\footnote{In this expression, $x^2+y^2$ is an implicit function of $Y$ and $Z$, following  \eqref{XYZxyz}.}
\begin{equation}
\label{Psiep0}
\text{e}^\Psi=x^2+y^2.
\end{equation}
The adapted Killings are $\mathscr{M}=\partial_\tau$ and $\mathscr{Z}=\partial_X$, whereas $\mathscr{X}$ and $\mathscr{Y}$ are more involved combinations of the new basis vectors.

In the case under investigation, both terms of \eqref{ds} are separately invariant under  $\mathscr{M}$ and $\mathscr{Z}$, whereas only their specific combination is invariant under $\mathscr{X}$ and $\mathscr{Y}$, completing the $\text{Heisenberg} \ltimes U(1)$ isometry algebra of this hyper-K\"ahler space. As a consequence, once we uplift this instanton to a quaternionic one using $\Psi(Y,Z)$ of \eqref{Psiep0} in Eqs.~\eqref{prztod},  \eqref{Aform} and \eqref{quatreq}, the resulting metric in no longer invariant under $\mathscr{X}$ and $\mathscr{Y}$. The Przanowski--Tod space has only $U(1)\times U(1)$ isometry, and belongs to the Calderbank--Pedersen class. It is not hard  to put its metric in the form \eqref{CPprztod} by trading $(t,x,y,z)$ for 
\begin{equation}
\label{coordep0}
\tau=\arctan \frac{x}{y}\,,\quad\psi=z+\frac{xy}{2}\,,\quad \rho=\sqrt{x^2+y^2}\,,\quad \eta=t\,,
\end{equation}
with 
\begin{equation}
\boxed{
\label{Gep0}
G=\frac{\eta\rho^2}{2}-\sigma\,.
}
\end{equation}
Here $\sigma$ is an arbitrary constant, showing that the quaternionic ascendent is rather a one-parameter family (this is actually a systematic  feature in all of our constructions).

To summarize, the quaternionic uplift of the unique $\text{Heisenberg}\ltimes U(1)$-symmetric hyper-K\"ahler space (Eq.~\eqref{hyp7} with $\varepsilon=0$) breaks the $\text{Heisenberg}\ltimes U(1)$ symmetry to $U(1)\times U(1)$. This space \emph{is not} the one found in \cite{oneloop}. The latter is a Calderbank--Pedersen space with 
\begin{equation}
\boxed{
\label{Goneloop}
G=\frac{\rho^2}{2}-\sigma\,,
}
\end{equation}
and extended  $\text{Heisenberg}\ltimes U(1)$ isometry. Its metric is explicitly displayed in Eq.  \eqref{CPM}.

\boldmath
\subsubsection*{The $\varepsilon=1$ uplift}
\unboldmath

We now turn to the hyper-K\"ahler space \eqref{hyp7} with $\varepsilon=1$. This provides a triaxial realization of the Heisenberg algebra, as no extra isometry appears.  
In order to express the metric in the Boyer--Finley form \eqref{ds}, \eqref{Toda.frame} and \eqref{Vom}, we must adapt the coordinate $\tau$ to $\mathscr{Y}$, which is now the available rotational Killing vector. This is achieved with the following change of coordinates
\begin{equation}
\label{chiy}
\tau=y\,,\quad
X=z\,,\quad Y= xt\,, \quad Z=\frac{1}{4}\left(\text{e}^{2t}(2t-1)-2x^2
\right)\,,
\end{equation}
while
\begin{equation}
\Psi=2t\,.
\end{equation}

Together with $\mathscr{Y}=\partial_\tau$, the Killing field $\mathscr{Z}$ is now adapted to the coordinate $X$ and  everything depends on $(Y,Z)$ only. The invariance under $\mathscr{X}$ (combination of $\partial_X,\partial_Y$ and $\partial_Z$) is the result of a fine cancellation  amongst the two terms in   \eqref{ds}. This cancellation no longer occurs in the uplifted quaternionic space, where $\omega$ is traded for  $\text{A}$ as in \eqref{Aform}, and $V$  for $U$ given in \eqref{quatreq}. Again, the 
Heisenberg symmetry is broken down to $U(1)\times U(1)$, generated by  $\mathscr{Y}$ and  $\mathscr{Z}$. In this case, the breaking was expected since no strict Heisenberg isometry exists at the quaternionic level, as shown in the beginning of the present  section.

The uplifted $\varepsilon=1$ quaternionic space is Calderbank--Pedersen, which we can put in the form \eqref{CPprztod} with 
\begin{equation}
\label{coordep1}
\tau=y\,,\quad\psi=z\,,\quad \rho=\text{e}^t\,,\quad \eta=x\, ,
\end{equation}
and
\begin{equation}
\boxed{
\label{Gep1}
G=\frac{1}{4}\left(
2\rho^2\ln\rho-2\eta^2-\rho^2\right)-\sigma\,,
}
\end{equation}
where $\sigma$ is an arbitrary constant.

\section{Heisenberg algebras and gaugings}\label{gaug}

To summarize at this point, any hyper-K\"ahler space with a rotational Killing vector can be uplifted to a quaternionic space, by going to the Boyer--Finley frame and using the available solution of the Toda equation, $\Psi(X,Y,Z)$. 
The original hyper-K\"ahler space appears as a vanishing-$k^2$ double-scaling limit of the quaternionic ascendent. 
In general however, this procedure does not respect the isometry content of the hyper-K\"ahler space
and there could be, as in the example of the Heisenberg$\ltimes U(1)$ isometry, another quaternionic ascendent
with identical isometry.
Using this construction, we have studied the quaternionic ascendents of the Heisenberg hyper-K\"ahler instantons \eqref{hyp7} with $\varepsilon=0$ and $\varepsilon=1$. We found in both cases Calderbank--Pedersen spaces with $U(1)\times U(1)$ isometry, and no further  extension. This is not surprising for $\varepsilon = 1$, as we have proven in Sec. \ref{genquater}  that no quaternionic space with triaxial Heisenberg symmetry exists. The isometry of the $\varepsilon = 1$ hyper-K\"ahler geometry could then only be broken in its quaternionic ascendent. 

The above results raise several questions, that we would like to investigate, at least partially, in the remaining of this note. Besides the interpretation of the two quaternionic spaces at hand in terms of string corrections to the hypermultiplet manifold, which seems out of reach at present, we would like to   
question the possibility of realizing these geometries in terms of gaugings inside the invariance group of the $\mathcal{N}=2$ hypermultiplet supergravity couplings,
 $Sp(2,4)$. We may wonder whether this is possible, and which generators should be gauged for reaching
 the two quaternionic ascendents we presented, Eqs.  \eqref{Gep0} and \eqref{Gep1}, as well as the already known $\text{Heisenberg}\ltimes U(1)$ space, Eq.~\eqref{Goneloop}. The way to proceed is to use hypermultiplet(s) coupled to local ${\cal N}=2$ superconformal symmetry \cite{deWit:1979ug,deWit:1984px} and to perform a quaternionic quotient \cite{Galicki:1985qv,Galicki:1987jz} using non-propagating vector multiplet(s) and additional hypermultiplet(s).

We will not present  an exhaustive analysis of all possible gaugings, but rather use this method in order to recover the quaternionic space missing in our previous approach of Sec. \ref{quater}: the one with  
$\text{Heisenberg}\ltimes U(1)$ symmetry discussed in \cite{oneloop} and displayed in Eq.~\eqref{Goneloop}. For this purpose, we need to gauge the generators $(\mathscr{Y},\mathscr{Z})$. We will not study here other concrete possibilities, as e.g.~the $(\mathscr{M},\mathscr{Z})$ gauging. We nevertheless wish to illustrate the power of the gauging procedure and use it to show that no triaxial Heisenberg symmetry could be reached for quaternionic spaces by gauging $SU(1,2)$ generators. This is achieved by analysing the little group of $SU(1,2)$ and demonstrating the absence of strict Heisenberg orbit. Since the general proof of non-existence of triaxial Heisenberg symmetry in quaternionic spaces was already presented in Sec. \ref{genquater},  we leave this complementary exercise for App. \ref{orbits.su12}.

The $(\mathscr{Y},\mathscr{Z})$ gauging provides a two-parameter family with generic  $U(1)\times U(1)$ generated by $\mathscr{Y}$ and $\mathscr{Z}$, which contains a one-parameter subfamily with $\text{Heisenberg}\ltimes U(1)$ \cite{AADT}. This family, is the one describing the one-loop perturbative corrections \cite{oneloop}. 
After a tedious computation, which is sketched in App.~\ref{gauging.appendix}, we find 
a  Calderbank--Pedersen space with 
\begin{equation}
\label{CPpotential.U1.U1}
\boxed{
G=\frac{\rho^2}{2}+\chi\,\eta-\sigma+2\chi^2\,,
}
\end{equation}
where $\chi$ and $\sigma$ are the two arbitrary parameters, associated respectively with
the $\mathscr{Y}$ and  $\mathscr{Z}$ gaugings.
The line element (Eqs. \eqref{CPprztod},  \eqref{CPprztod3d} and  \eqref{CPgeneral1}) reads:
\begin{eqnarray}
\mathrm{d}s^2&=&\frac{2V_1}{k^2V_2^2}\,
\left(\mathrm{d}\rho^2+\mathrm{d}\eta^2+\frac{\rho^2}{\rho^2+\chi^2}\,\mathrm{d}\psi^2\right) \nonumber \\
&&+\frac{8\left(\rho^2+\chi\right)}{k^2V_1V_2^2}\,\left(\mathrm{d}\tau+
\frac{2\eta\, \rho^2-\chi\left(4\chi^2
-2\sigma+\rho^2
\right)}{2\left(\rho^2+\chi^2\right)}\mathrm{d}\psi\right)^2
\label{CPmetric.U1.U1}
\end{eqnarray}
with
\begin{equation}
\label{CPmetric.U1.U1.2}
V_1=\rho^2+2\left(\sigma-\chi(\eta+\chi)\right)\,,\quad V_2=\rho^2-2\left(\sigma-\chi(\eta+2\chi)\right)\,.
\end{equation}

The family of Calderbank--Pedersen spaces at hand possesses generically two commuting isometries generated by  $\mathscr{Y}=\partial_\psi$ and $\mathscr{Z}=\partial_\tau$. For vanishing $\chi$, the function $G(\rho, \eta)$ in 
Eq.~\eqref{CPpotential.U1.U1} matches that of Eq.~\eqref{Goneloop}, whereas the metric \eqref{CPmetric.U1.U1}, coincides with \eqref{CPM}. This is where the symmetry is extended to $\text{Heisenberg}\ltimes U(1)$, with two extra Killing fields $\mathscr{X}, \mathscr{M}$. When $\sigma$ also vanishes, the isometry is further enhanced to $U(1,2)$, and the metric is the non-compact Fubini--Study. All these properties are analysed in App.~\ref{gauging.appendix} from the gauging perspective. 

Our next task is to find a convenient zooming-in limit when $k\to0$, leading to a hyper-K\"ahler space for the family  \eqref{CPmetric.U1.U1}.
Such a limit requires the Kretschmann 
($K=R_{\kappa\lambda\mu\nu}R^{\kappa\lambda\mu\nu}$) scalar
to remain finite. In this limit, however, the latter vanishes unless
\begin{equation}
 V_1\to 0\,.
\end{equation}
In order for the line element to remain regular when $V_1$ vanishes, we should perform the appropriate coordinate rescalings, keeping in particular the $\tau$-fiber finite (see also \cite{AADT}). We define new coordinates $t,x,y,z$ as 
\begin{equation}
\begin{split}
&\rho^2=4\chi^2-2\sigma+(t+2\chi x)\left(\frac{k}{1+4\chi^2}\right)^{\nicefrac{2}{3}}
\,, \quad
 \psi=y
\frac{k^{\nicefrac{2}{3}}}{\left(1+4\chi^2\right)^{\nicefrac{1}{6}}}
\,, \\
&
\eta=\chi+(x-2\chi t)\left(\frac{k}{1+4\chi^2}\right)^{\nicefrac{2}{3}}
\,,\quad
 \tau=z\, k^{\nicefrac{4}{3}}\left(1+4\chi^2\right)^{\nicefrac{1}{6}}
\,.
\end{split}
\end{equation}
Hence, the limit $k\to 0$ amounts to zooming around the point\footnote{This assumes $2\chi^2\geqslant\sigma$, otherwise the limit would reproduce Euclidean flat  space.}  $(\rho^2_0,\eta_0)=(4\chi^2-2\sigma,\chi)$, 
and obtain
\begin{equation}
\label{HKlimit}
\mathrm{d}s^2_{k\to0}=\frac{1}{t}\left(\mathrm{d}z+x\mathrm{d}y\right)^2+
t\,\left(\mathrm{d}t^2+\mathrm{d}x^2+\mathrm{d}y^2\right)\,,
\end{equation}
This is the hyper-K\"ahler metric with $\text{Heisenberg} \ltimes U(1)$, given in  Eq. \eqref{hyp7} with $\varepsilon=0$.

In conclusion, \textit{the two-parameter gauging of $(\mathscr{Y},\mathscr{Z})$  generators leads to a quaternionic manifold with $U(1)\times U(1)$
isometry (Eq.~\eqref{CPmetric.U1.U1}), which, in the flat limit, reproduces the hyper-K\"ahler metric with $\text{Heisenberg} \ltimes U(1)$ symmetry (Eq.~\eqref{hyp7} with $\varepsilon=0$).}

\section*{Conclusions and outlook}
\addcontentsline{toc}{section}{Conclusions}

An important result
in the present note is the obstruction for a quaternionic space to host a strict (triaxial) Heisenberg algebra. 
This can be rigorously demonstrated in at least two ways, we have chosen to 
use the foliation technique in Sec. \ref{genquater} and further showed
in App.~\ref{orbits.su12} how the gauging techniques operate in the same direction (excluding the strict Heisenberg orbit). 

The above obstruction is illustrated when scanning over the landscape of hyper-K\"ahler and quaternionic spaces. As a tool for such a scanning, we introduced a method which allows to uplift hyper-K\"ahler geometries possessing rotational Killing vectors, to quaternionic spaces. Indeed, both hyper-K\"ahler spaces with a rotational symmetry and  quaternionic spaces with a symmetry rely on a solution of the continual Toda equation, and this solution bridges the two geometries  (conversely, a systematic $k\to 0$ zooming-in in the quaternionic space enables us to recover the original  hyper-K\"ahler geometry, and this is useful when the latter appears in the global $\mathcal{N}=2$ limit). 
In the course of the uplifting, part of the extra symmetries are usually lost. Applied to either of the two Bianchi II hyper-K\"ahler spaces, \emph{i.e.} with biaxial ($\text{Heisenberg}\ltimes U(1)$), or triaxial (strict Heisenberg) realization of the symmetry, the proposed uplift leads to a Calderbank--Pedersen geometry with only $U(1)\times U(1)$ isometry.

In order to make contact with supergravity applications, it is useful to recover the  
quaternionic space known to capture the perturbative corrections in the type IIA hypermultiplet scalar manifold, Eq. \eqref{Goneloop}. The global $\mathcal{N}=2$ limit of this space is again the unique biaxial 
$\text{Heisenberg}\ltimes U(1)$ hyper-K\"ahler. However, the sought for quaternionic space is not obtained using the above uplifting procedure. Instead, performing a gauging in the appropriate directions inside the $Sp(2,4)$ algebra, a full family of Calderbank--Pedersen spaces with $U(1)\times U(1)$ symmetry is reached, 
Eq.~\eqref{CPpotential.U1.U1}, which contains an extended-symmetry point where 
$\text{Heisenberg}\ltimes U(1)$ is realized. This is the corrected type IIA hypermultiplet scalar manifold. 

The above analysis is summarized in Tab. \ref{landscape}. Regarding gaugings, a question is raised that remains open at the present stage of our discussion: can one design gaugings that would reproduce the two Calderbank--Pedersen spaces reported in Eqs. \eqref{Gep0} and \eqref{Gep1}, whose  $k\to 0$ limit are the Bianchi II 
hyper-K\"ahler instantons (last two lines of Tab. \ref{landscape})? This question is naturally accompanied by a more physical one: do these 
spaces, related to the Heisenberg algebra, admit any supergravity interpretation in connection with the hypermultiplet scalar manifold? We plan to come back to these issues in the future.

\begin{table}[h]
\begin{center}
    \begin{tabular}{| l | l | l || l |}
    \hline
    Quaternionic construction & $G(\rho,\eta)$ & Symmetry & Flat limit \\ \hline\hline
    Hyper-K\"ahler cone \cite{Anguelova:2004sj}
    & $\nicefrac{\rho^2}{2}-\sigma$  & $\text{Heisenberg}\ltimes U(1)$ & $\text{HK}_{\varepsilon=0}$ \\ 
 \hline
   $\mathscr{Z}$ gauging \cite{AADT}
   & $\nicefrac{\rho^2}{2}-\sigma$  & $\text{Heisenberg}\ltimes U(1)$ &  $\text{HK}_{\varepsilon=0}$ \\ \hline
    $(\mathscr{Y},\mathscr{Z})$ gauging & $\nicefrac{\rho^2}{2}+\chi\,\eta-\sigma+2\chi^2$  & $U(1)\times U(1)$ &  $\text{HK}_{\varepsilon=0}$ \\ \hline
    Uplift of the $\text{HK}_{\varepsilon=0}$ & $\eta\rho^2-\sigma$  & $U(1)\times U(1)$ &  $\text{HK}_{\varepsilon=0}$\\
   \hline
    Uplift of the $\text{HK}_{\varepsilon=1}$ & $\nicefrac{1}{4}\left(
2\rho^2\ln\rho-2\eta^2-\rho^2\right)-\sigma$ & $U(1)\times U(1)$ &  $\text{HK}_{\varepsilon=1}$ \\ 
\hline
    \end{tabular}
\end{center}
\vskip-0.6cm
\caption{Summary of the various quaternionic geometries, their origins, symmetries  and flat limits -- $\text{HK}_{\varepsilon}$ stands for the hyper-K\"ahler family \eqref{hyp1} with $\text{Heisenberg}\ltimes U(1)$ ($\varepsilon=0$), or strict Heisenberg symmetry ($\varepsilon=1$).}
    \label{landscape}   
\end{table}

\section*{Acknowledgements}

We would like to thank Liana David and Paul Gauduchon for useful correspondence, Pierre Vanhove for a
fruitful conversation and Konstantinos Sfetsos for priceless 
exchange on the continual Toda and related problems.
Part of this work was developed during HEP 2015 -- \textsl{Conference on Recent Developments in High Energy Physics and Cosmology} in April 2015 at the National and Kapodistrian University of Athens.
 We also acknowledge each-other institutes for hospitality and financial support. This work was partially supported by the \textsl{Germaine de Stael} Franco--Swiss bilateral program 2015 (project no 32753SG).

\appendix

\boldmath
\section{Gaugings}
\unboldmath
\label{gauging.appendix}

The purview of this appendix is to provide the technical details of Sec. 
\ref{gaug} on the $(\mathscr{Y},\mathscr{Z})$ gauging.
Our discussion will closely follow the $\mathscr{Z}$ gauging performed in Sec. 3 of  \cite{AADT}.

Let us consider three hypermultiplets coupled to ${\cal N}=2$ superconformal supergravity.
The physical hypermultiplet has positive signature, whereas the compensating ones and the non-propagating vector have
negative signature. The hypermultiplet scalars are $A_i^\alpha$, with $SU(2)_R$ index $i=1,2$ and $Sp(2,4)$ index 
$\alpha=1,\ldots,6$. They transform in the representation $({\bf6},{\bf2})$ of $Sp(2,4)\times SU(2)_R$.
Their conjugates are
$A^i_\alpha = (A_i^\alpha)^* = \varepsilon^{ij}\rho_{\alpha\beta} A^\beta_j$
with $\rho^{\alpha\beta}\rho_{\beta\gamma} = - \delta^\alpha_\gamma$ and 
$\varepsilon^{ij}\varepsilon_{jk} = - \delta^i_k$\,. 
We choose the $Sp(2,4)$-invariant metric as 
\begin{equation}
\label{conf3}
\rho =   i\,\sigma_2\otimes \mathbb{I}_3 =
\begin{pmatrix} 
 0 & \mathbb{I}_3  \\ -\mathbb{I}_3  & 0 
 \end{pmatrix}
\end{equation}
and\footnote{Our choice of $\eta$ incorporates the nature of the hypermultiplets, such that the constraints can be solved \cite{AADT}.} 
\begin{equation}
\label{eta.appendixA}
d = \begin{pmatrix} \eta&0\\0&\eta\end{pmatrix},
\quad\eta = {\rm diag}( -1 , 1 , -1 ), \quad
\rho \,d\, \rho = -d.
\end{equation}

At the tree-level, the universal dilaton hypermultiplet is mapped, after Poincar\'e duality, to
the quaternionic and K\"ahler pseudo-Fubini--Study metric of the coset space $\frac{SU(1,2)}{U(2)}$ \cite{Dancer.Einstein}.
At one-loop, the isometry is lessened to the Heisenberg subalgebra of $SU(1,2)$, which is generated by the following 
three elements:
\begin{equation}
\label{Heisenberg.matrix}
\mathscr{X} = \begin{pmatrix} 0 & 0 & 1 \\ 0 & 0 & 1 \\ -1 & 1 & 0  \end{pmatrix},\quad
\mathscr{Y} = \begin{pmatrix} 0 & 0 & i \\ 0 & 0 & i \\ i & -i & 0  \end{pmatrix},\quad
\mathscr{Z} = 2\begin{pmatrix} i & -i & 0 \\ i & -i & 0 \\0 & 0 & 0  \end{pmatrix},\quad
\end{equation}
with $\left[\mathscr{X},\mathscr{Y}\right]=\mathscr{Z}$.
We will gauge 
\begin{equation}
\label{gauging.T}
\hat T = i\, \mathbb{I}_3 + \chi\, \mathscr{Y} + \sigma \mathscr{Z}\,,
\end{equation}
where $(\chi,\sigma)$ are two arbitrary parameters. 
This gauging contains a one-parameter subfamily with $\text{Heisenberg}\ltimes U(1)$, as $\hat T$
commutes with it, 
where the $U(1)$ is generated by
\begin{equation}
\mathscr{M} = \frac{i}{3}\begin{pmatrix}1 & 0 & 0 \\ 0 & 1 & 0 \\ 0 & 0 & -2  \end{pmatrix},
\end{equation}
and obeys $\left[\mathscr{M},\mathscr{X}\right]=\mathscr{Y}$ and $\left[\mathscr{M},\mathscr{Y}\right]=-\mathscr{X}.$

We now come to the ${\cal N}=2$ conformal supergravity Lagrangian.
Eliminating the auxiliary fields from the gauge-fixing of the dilation symmetry in the Poincar\'e  theory, we find~\cite{AADT}: 
\begin{equation}
\begin{split}
\label{LagrangianN2}
e^{-1}{\cal L}=\text{Tr}(\partial_\mu A^\dagger)d(\partial^\mu A)-g'^2\text{Tr}(A^\dagger T^\dagger d T A) W^\mu W_\mu-
\frac{g^2}{k^2}\text{Tr} (V^\mu V_\mu)\,,\\
W_\mu=\frac{ \text{Tr}(\partial_\mu A^\dagger d TA-A^\dagger d T\partial_\mu A)}{2g'{\text{Tr}}(A^\dagger T^\dagger d T A)}\,,\quad
V_\mu=-\frac{\partial_\mu A^\dagger d A-A^\dagger d\partial_\mu A}{g\,{\text{Tr}}(A^\dagger d A)}\,.
\end{split}
\end{equation}
Here $A$ is a complex-doublet vector of components $A^\alpha_i$ (and $A_\alpha^i$ for its complex-conjugate)
\begin{equation}
A = \begin{pmatrix}  \vec A_+ & \vec A_- \\
-\vec A_-^* & \vec A_+^*   \end{pmatrix},  \quad
A^* = \begin{pmatrix} \vec A_+^* & \vec A_-^* \\
-\vec A_- & \vec A_+   \end{pmatrix},
\end{equation}
subject to the constraints 
\begin{equation}
{\text{Tr}}A^\dagger d A=-\frac{2}{k^2}\,,\quad A^\dagger d T A=0\,,
\end{equation}
and $T$ the $Sp(2,4)$ extension of \eqref{gauging.T}:
\begin{equation}
T =  \begin{pmatrix} \hat T & 0  \\  0 & \hat T^* \end{pmatrix} \,.
\end{equation}

Working along the same lines of \cite{AADT} we find the solution for the constraints:
\begin{equation}
\label{Asolution}
\vec A_+ = \frac{1}{\Delta} \begin{pmatrix} 2S - (\Phi+\chi)^2  - 1 \\
2S - (\Phi+\chi)^2 + 1 \\ 2 (\Phi + \chi)
\end{pmatrix}, \quad
\vec A_- =  \frac{K}{\Delta}\begin{pmatrix} \ov\Phi \\ \ov \Phi \\ 1  \end{pmatrix}\,,
\end{equation}
where $K,\Delta$ are real quantities given by
\begin{equation}
\begin{split}
&K^2 = -4\left(S+\ov S\right) + 2\left(\Phi+\ov\Phi\right)\left(\Phi+\ov\Phi+2\chi\right) + 4\sigma\,,\\
&\Delta^2 = -4\left(S+\ov S\right) + 2\left(\Phi+\ov\Phi+\chi\right)\left(\Phi+\ov\Phi+2\chi\right) + 2\sigma\,.
\end{split}
\end{equation}
The scalar kinetic Lagrangian \eqref{LagrangianN2} obtained with \eqref{Asolution} reads:
\begin{equation}
k^2{\cal L}={\cal I}_1+{\cal I}_2+{\cal I}_3\,,
\end{equation}
with
\begin{equation}
\label{general.sigma}
\begin{split}
&{\cal I}_1=\frac{1}{\Delta^2}\left(2(\partial\Delta)^2-
\frac{\left(\Delta\partial\Delta-2\chi{\rm Re}\,\partial\Phi\right)^2}{\Delta^2-4\chi{\rm Re}\,\Phi+2\sigma-4\chi^2}
-4\left\|\partial\Phi\right\|^2\right)\,,\\
&{\cal I}_2=\frac{8\left[{\rm Im}\left(\partial S-(2{\rm Re}\,\Phi+\chi)\,\partial\Phi\right)\right]^2}
{\Delta^2\left(\Delta^2-8\chi\,{\rm Re}\,\Phi+4\sigma-6\chi^2\right)}\,,\\
&{\cal I}_3=\frac{8\left[{\rm Im}\left(\partial S-2({\rm Re}\,\Phi+\chi)\,\partial\Phi\right)
\right]^2}{\Delta^4}
+\frac{2}{\Delta^4}\left\|\chi\partial K-2K\partial\Phi\right\|^2\,,
\end{split}
\end{equation}
where $\left\|A\right\|^2=A_\mu \ov A^\mu.$
The Lagrangian can be expressed as usual:
\begin{equation}
{\cal L}=\frac{1}{2k^2}\,g_{ab}\,\partial_\mu q^a\,\partial^\mu q^b\,,
\end{equation}
where $\mathbf{q}=(\Delta^2,{\rm Re}\,\Phi,{\rm Im}\,\Phi,{\rm Im} S)$, and we introduce  
$\tau, \eta$ and $\psi $ as $\tau={\rm Im}S$ and $\nicefrac\eta2+i\psi=\Phi$. 
The metric $G_{ab}=\nicefrac{g_{ab}}{k^2}$ describes a Calderbank--Pedersen space with
\begin{equation}
\label{CPmetric.U1.U1potential}
G(\rho,\eta)=\frac{\rho^2}{2}+\chi\,\eta-\sigma+2\chi^2\,,
\end{equation}
where 
$\rho=\sqrt{\Delta^2-2\chi\,\eta+2\sigma-4\chi^2}\,.$ 
The corresponding
line element reads:
\begin{eqnarray}
\mathrm{d}s^2&=&\frac{2V_1}{k^2V_2^2}\,
\left(\mathrm{d}\rho^2+\mathrm{d}\eta^2+\frac{\rho^2}{\rho^2+\chi^2}\,\mathrm{d}\psi^2\right) \nonumber \\
&&+\frac{8\left(\rho^2+\chi\right)}{k^2V_1V_2^2}\,\left(\mathrm{d}\tau+
\frac{2\eta\, \rho^2-\chi\left(\rho^2-2\sigma+4\chi^2\right)}{2\left(\rho^2+\chi^2\right)}\mathrm{d}\psi\right)^2
\label{CPmetric.U1.U1appendix}
 \nonumber
\end{eqnarray}
with
$V_1=\rho^2+2\left(\sigma-\chi(\eta+\chi)\right)$ and $V_2=\rho^2-2\left(\sigma-\chi(\eta+2\chi)\right)$,
as reported already in Eqs. \eqref{CPmetric.U1.U1} and  \eqref{CPmetric.U1.U1.2}.

\boldmath
\section{Orbits of $SU(1,2)$ generators}
\unboldmath
\label{orbits.su12}

The scope of this appendix is to determine the little group of the $SU(1,2)$ adjoint representation
and to show that there is no strict Heisenberg orbit. This demonstrates the absence of quaternionic 
spaces with triaxial Heisenberg isometry, which would have been obtained by gauging 
 $SU(1,2)$ generators in the spirit of Sec.  
\ref{gaug} and App. \ref{gauging.appendix}. It also provides an alternative perspective to the incompatibility of strict Heisenberg symmetry with quaternionic spaces, proven in Sec. \ref{genquater}. 

First, consider the compact case.
Any (antihermitian) generator of $SU(3)$ for the three-dimensional representation can be diagonalized
with imaginary eigenvalue $a$, $b$, $c$ verifying $a+b+c=0$. The little group is  
either $SU(2)\times U(1)$, which is four-dimensional, if two eigenvalues are equal 
or $U(1)\times U(1)$ (Cartan algebra, two-dimensional) if all three eigenvalues differ. Hence
the orbits of the adjoint representation have two- and four-dimensional stability group (besides of course 
the trivial orbit with $SU(3)$).
In the non-compact $SU(1,2)$ case, this result remains true, but the identification of the little groups
is more subtle.

Let $\mathscr{U}$ be an element of the $\frak{su}(1,2)$ algebra with invariant metric $\eta=\text{diag}(-1,1,1)$. The first
direction is then time-like.\footnote{Notice that $\eta$ differs from the non-standard
choice used in Eq. \eqref{eta.appendixA}.}
We shall parameterize $\mathscr{U}$ as 
\begin{equation}
\label{elementU}
\mathscr{U}=\left(\begin{matrix} 
i a & A & B \\
\ov A& i b &C\\
\ov B & -\ov C & i c
\end{matrix}\right)\,,\quad a+b+c=0\,,\quad (a,b,c)\in\mathbb{R}\,,\quad (A,B,C)\in\mathbb{C}.
\end{equation}
The non-compact $SU(1,2)$ has four Jordan conjugacy classes \cite{Gauduchon}. Following the 
terminology of Ref.  \cite{Gauduchon}, we shall refer to them as {\it elliptic}, {\it hyperbolic}, {\it one-step parabolic} and {\it two-step parabolic}. 
\begin{enumerate}
\item 
In the {\it elliptic} class, $\mathscr{U}$ admits a time-like eigenvector with imaginary 
eigenvalue. One can then choose $A=B=0$ and further diagonalize the remaining compact directions ($C=0$).
The little groups are then obviously:
\begin{equation}
{\cal G}(\mathscr{U})=
	\begin{cases}
		U(1)\times U(1)\,,  & a\neq b\neq c\neq a \,,\\
		SU(1,1)\times U(1)\,, & a=b \quad{\rm or}\quad a=c\,,\\
		SU(2)\times U(1)\,, & b=c\,,\\
		SU(1,2)\,, & a=b=c=0\,.
	\end{cases}
\end{equation}
In the second case, the $U(1)$ is generated by $\mathscr{M}$ (which is  present in all four cases).
\item 
In the {\it hyperbolic} class, $\mathscr{U}$ has a space-like eigenvector with imaginary eigenvalue ($C=B=0$)
and null (light-like) eigenvectors (non-orthogonal in metric $\eta$, with eigenvalues $A$ and $-\ov A$). It takes the form:
\begin{equation}
\mathscr{U}=\begin{pmatrix} 
i\,{\rm Im} \,A & {\rm Re} \,A &0 \\
{\rm Re} \,A & i\,{\rm Im} \,A &0\\
0 & 0 & -2i\,{\rm Im} \,A
\end{pmatrix}.
\end{equation}
It is a linear combination of the two commuting generators of the little group:
${\cal G}(\mathscr{U})=U(1)\times U(1)$. One generator is $\mathscr{M}$, the second is in the $SU(1,1)$,
which commutes with~$\mathscr{M}$.
\item
In the {\it one-step parabolic} class, $\mathscr{U}$ can be written as:
\begin{equation}
\mathscr{U}=i\begin{pmatrix} 
\lambda+a & - a \text{e}^{-i\varphi} &0 \\
a \text{e}^{i\varphi}  & \lambda-a &0\\
0 & 0 & -2\lambda
\end{pmatrix}\,
\end{equation}
with $(\lambda,a,\varphi)\in\mathbb{R}$. It admits one space-like and one light-like eigenvector with 
imaginary eigenvalue $i\lambda$.
\begin{itemize}
\item For $\lambda\neq0$, we find the little group  ${\cal G}(\mathscr{U}) = U(1)\times U(1)$, as in case 2.
\item For $\lambda=0$, one can write
\begin{equation}
\mathscr{U}= ia \begin{pmatrix} 
1 & -\text{e}^{-i\varphi} &0 \\
\text{e}^{i\varphi}  & -1 &0\\
0 & 0 & 0
\end{pmatrix}
= \frac{a}{2} \mathscr{R}^\dagger \, \mathscr{Z} \, \mathscr{R}, 
\qquad
\mathscr{R} = \begin{pmatrix} 
\text{e}^{i\nicefrac{\varphi}{2}} & 0 &0 \\
0 & \text{e}^{-i\nicefrac{\varphi}{2}}  &0 \\
0 & 0 & 1
\end{pmatrix}
\end{equation}
and, since $\mathscr{R}$ is an $SU(1,2)$ element, $\mathscr{U}$ is in the same orbit as $-\frac{a}{2}\mathscr{Z}$
which is known to commute, as one can explicitly verify, with the four-dimensional Heisen\-berg$\,\ltimes U(1)$ algebra.

\end{itemize}
\item
In the {\it two-step parabolic} class, 
\begin{equation}
\mathscr{U}= \mathscr{Y}=i\begin{pmatrix} 
0 & 0 &1 \\
0 & 0 &1\\
-1 & 1 & 0
\end{pmatrix}
\end{equation}
with triple zero eigenvalue and a light-like eigenvector. 
The little group is ${\cal G}(\mathscr{U})=U(1)\times U(1)$, generated by
$\mathscr{Y}$ and $\mathscr{Z}$.

\end{enumerate}
We can recapitulate the above results, regarding the possible gaugings within $SU(1,2)$, as follows. 
\textit{The little groups of the $SU(1,2)$ adjoint representation can be either: 
\begin{itemize}
\item
2-dimensional: $U(1)\times U(1)$;
\item
4-dimensional:
$SU(2)\times U(1)$, or $SU(1,1)\times U(1)$, or $\text{Heisenberg} \ltimes U(1)$;
\item
8-dimensional: $SU(1,2)$.
\end{itemize}
There is no three-dimensional little group in $SU(1,2)$, hence the option of strict Heisenberg symmetry is not available.}


\end{document}